\documentclass{aa}
\usepackage{graphicx,natbib}
\bibpunct{(}{)}{;}{a}{}{,}

\def\sgrb{Sgr~B2}
\def\sgra{Sgr~A$^{\star}$}
\def\igr{IGR~J17475$-$2822} 
\def\taut{\tau_{\rm T}} 
\def\sigmat{\sigma_{\rm T}} 
\def\mpr{m_{\rm p}} 
\def\ms{M_\odot} 

\begin{document}

\title{Hard X-ray view of the past activity of \sgra\ in a natural Compton
mirror}  

\author{
M.G. Revnivtsev\inst{1,2},
E.M. Churazov\inst{1,2},
S.Yu. Sazonov\inst{1,2},
R.A. Sunyaev\inst{1,2},
A.A. Lutovinov\inst{1},\\
M.R. Gilfanov\inst{1,2},
A.A. Vikhlinin\inst{1,3},
P.E. Shtykovsky\inst{1} and 
M.N. Pavlinsky\inst{1}
}

\offprints{revnivtsev@hea.iki.rssi.ru}

\institute{Space Research Institute, Russian
Academy of Sciences, Profsoyuznaya 84/32, 117997 Moscow, Russia
\and 
Max-Planck-Institut f\"ur Astrophysik,
Karl-Schwarzschild-Str. 1, D-85740 Garching bei M\"unchen, Germany
\and 
Harvard-Smithsonian Center for Astrophysics, 60
Garden Street, Cambridge, MA 02138, USA
}
\date{\today}

\titlerunning{Hard X-ray view of past activity of \sgra\ in a natural
Compton mirror}
\authorrunning{Revnivtsev et al.}

\abstract{We report the association of the recently discovered hard
X-ray source \igr\ with the giant molecular cloud \sgrb\ in
the Galactic Center region. The broad band (3--200\,keV) spectrum of
the source constructed from data of different observatories strongly supports
the idea that the X-ray emission of \sgrb\ is Compton scattered
and reprocessed radiation emitted in the past by the \sgra\ source. We
conclude that 300--400 years ago \sgra\ was a low luminosity ($L\approx
1.5\times 10^{39}$\,erg\,s$^{-1}$ at 2--200\,keV) AGN with a
characteristic hard X-ray spectrum (photon index $\Gamma\approx
1.8$). We estimate the mass and iron abundance of the \sgrb\
scattering gas at $2\times 10^{6}M_\odot(r/10\,{\rm pc})^2$ and
1.9 solar, respectively (where $r$ is the radius of the
cloud).  
}
\maketitle

\section{Introduction}
Our Galactic Center (GC) harbors a black hole (BH) with a mass of
$3\times 10^6\ms$ (e.g. \citealt{sog+03}). It has remained
puzzling why the source \sgra\ associated with the BH is faint despite
the presence of significant amounts of ambient gas capable of fuelling it
\citep{mf01,bmm+03}. 

Among many complex structures near the GC, X-ray
telescopes have detected 8--20\,keV continuum \citep{smp93,msp93} and
6.4\,keV line diffuse emission \citep{kms+96,mks+00,smt+01,mkm01} 
associated with giant molecular clouds, in particular \sgrb\ located
at a projected distance of $\sim$100\,pc from \sgra. That was
suggested to be radiation emitted in the past by \sgra, Compton
scattered and reprocessed by the cloud neutral gas and delayed by
the light travel time \citep{smp93,kms+96}. 

The scattered emission is strongly photoabsorbed within the \sgrb\
cloud at energies below 5--10\,keV. However, since the efficiency of
photoabsorption rapidly declines with energy, one could expect \sgrb\
to be a strong X-ray source at energies above $\sim$15\,keV. We show
below that such a hard X-ray source has now been observed with the
INTEGRAL observatory. 

\section{Observations and results}
The INTEGRAL satellite  \citep{wcd+03} expands the frontiers of X-ray
and gamma-ray imaging. The several-fold increased effective area of
the telescope IBIS \citep{uld+03} compared to the previous
space-borne hard X-ray imager GRANAT/SIGMA, opens up a
possibility to study hard X-ray sources located in crowded regions 
of the sky, at flux levels (above 20\,keV) down to 1 mCrab \citep{rsv+04}.    

During the period 2003--2004 INTEGRAL extensively observed the
central $20^\circ\times 20^\circ$ area of the Galaxy
\citep{rsv+04}. Figure~\ref{fig:map} shows a hard X-ray image of the
innermost $3.5^\circ\times 2.5^\circ$ region obtained from
observations with a total effective exposure of 2.3\,Ms. Thanks to the
good angular resolution ($12^\prime$) of IBIS, practically all bright
sources in this area are resolved, most of them being known low-mass
X-ray binaries. The newly discovered source \igr\ \citep{rsv+04} is
coincident with the \sgrb\ molecular cloud. Since the X-ray flux below
10\,keV collected from the IBIS PSF centered on \igr\ is completely
dominated by diffuse emission of \sgrb\ \citep{smt+01,mkm01}, we can
safely associate \igr\ with this diffuse component. The observed flux in the 
20--200\,keV band is $2.5\pm0.1$~mCrab, or $\sim 7\times 
10^{-11}$\,erg\,s$^{-1}$, which corresponds to a luminosity of
$6\times10^{35}$\,erg\,s$^{-1}$ at a distance of 8.5\,kpc. To our
knowledge, this is the first ever detection of X-ray emission above
20\,keV from any molecular cloud.

We constructed a broad band spectrum of the source
(Fig.~\ref{fig:spectrum}) by combining data of GRANAT, ASCA and
INTEGRAL. The ASCA/GIS spectrum was collected from a 
$6.5^\prime$-radius region centered on \igr\, using data of
observations carried out in 1993--1994. The GRANAT/ART-P spectral
point was obtained from observations in 1991--1993 \citep{pgs94}. Note
that the different effective beam sizes of the instruments
($\sim$5$^\prime$ for GRANAT/ART-P and $\sim$12$^\prime$ for
INTEGRAL/IBIS) can affect the obtained spectrum. This effect should
however be small given the limited size (3--5$^\prime$) of the
Sgr B2 cloud in X-rays.

\section{Discussion}

\subsection{X-ray reflection nebula model}

The 2--10\,keV flux from \sgrb\ is dominated by diffuse 
emission in a line at 6.4\,keV superposed on strongly absorbed
continuum emission \citep{kms+96}. A number of smaller molecular clouds
in the GC region also exhibit powerful 6.4\,keV line emission,
although with low absolute fluxes compared to \sgrb\
\citep{bms+02,pch+03,pmb+04}. Common for all of these sources is the huge
(1--2\,keV) equivalent width of the 6.4\,keV line. 

\begin{figure}
\begin{center}
\includegraphics[width=0.8\columnwidth]{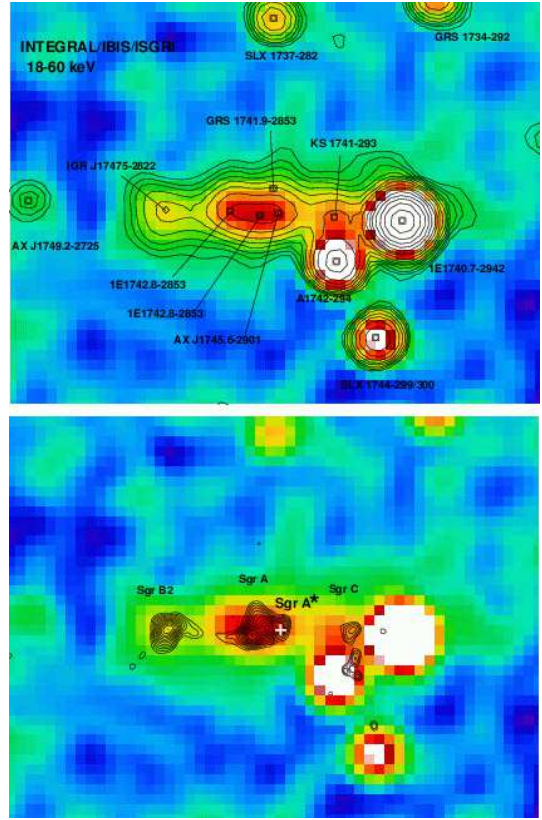}
\end{center}
\caption{Upper panel: $3.5^\circ\times 2.5^\circ$ hard X-ray
(18--60\,keV) image of the GC region obtained with
INTEGRAL/IBIS. Contours denote levels of the signal to noise ratio,
which start from S/N=5.0 and increase with a multiplicative 
factor of 1.4. Detected known X-ray sources are indicated (see
\citealt{rsv+04} for details). Lower panel: The same IBIS color image,
with overplotted contours of brightness distribution in the 6.4\,keV
line as measured by ASCA/GIS. Largest molecular clouds are indicated
and the position of the \sgra\ source is marked with a cross.} 
\label{fig:map}
\end{figure}

According to an early prediction \citep{smp93}, X-ray activity of
\sgra\ in the recent past should lead just to such observational
consequences, i.e. to the appearance of scattered X-ray radiation,
strongly photoabsorbed at low energies, and of a powerful fluorescent
iron K$_\alpha$ line. The X-ray echo from \sgrb\ should be delayed by
300--400 years relative to the direct signal from \sgra\ due to the
light travel time from \sgra\ to \sgrb. The equivalent width of the
K$_\alpha$ line is so large because we do not see the primary source
itself. The fact that INTEGRAL sees X-ray emission above 20\,keV from
the zone of 6.4\,keV emission in \sgrb\ (see Figure~\ref{fig:map})
provides strong support to this scenario. Unfortunately, the angular
resolution of IBIS is not sufficient to study the diffuse emission of
the other GC molecular clouds due to the presence of a large number of
strong point sources.

\begin{figure}
\includegraphics[width=\columnwidth,bb=25 182 567 698,clip]{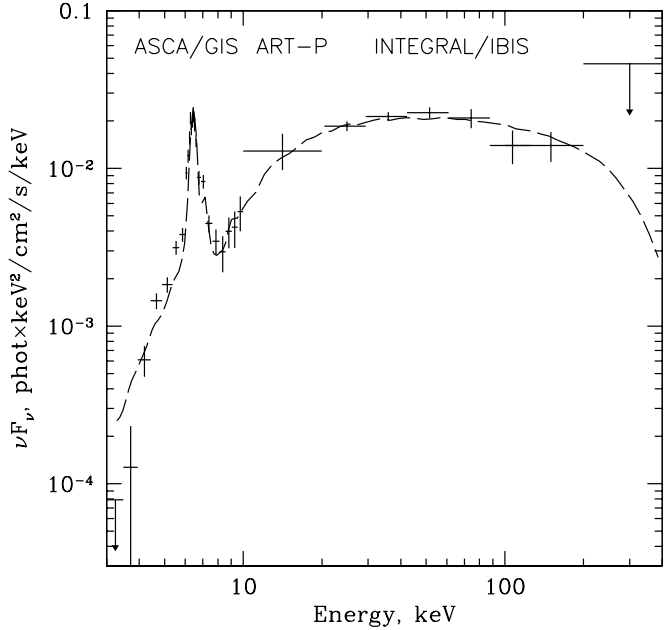} 
\caption{Broad band X-ray spectrum of the source \igr\ associated with the
\sgrb\ cloud. Data of ASCA/GIS (3--10\,keV), GRANAT/ART-P 
(10--20\,keV) and INTEGRAL/IBIS (20--400\,keV) are presented. $1\sigma$ error
bars and $2\sigma$ upper limits are shown. The dashed line is the
best-fit model (see main text) convolved with the resolution of
ASCA/GIS ($\sigma\approx 230$\,eV).}
\label{fig:spectrum}
\end{figure}

Our spectral analysis further supports the \sgra\ scenario. As shown
in Fig.~\ref{fig:spectrum}, the spectral energy distribution of
\sgrb\ measured with INTEGRAL at 20--200\,keV matches the 3--20\,keV
spectrum measured with ASCA and GRANAT/ART-P. The combined spectrum
at 3--200\,keV can be well fit by a model in which X-rays from
\sgra\ are scattered and reprocessed in a homogeneous spherical cloud
of cold gas. Scattering of the hard X-rays occurs on the neutral molecular
hydrogen and helium \citep{sc96} while the abundance of iron determines the 
intensity of the fluorescent K$_\alpha$ line. 

The spectrum emerging from \sgrb\ depends on several parameters: the
slope ($\Gamma$) of the incident spectrum (assumed to be a power law),
the cloud radial optical depth to Thomson scattering
[$\taut=\sigmat(2n_{{\rm H}_2})r$, where $n_{{\rm H}_2}$ is the number
density of hydrogen molecules and $r$ is the cloud radius], the iron
abundance relative to solar ($A$), the scattering angle ($\theta$)
for photons travelling from \sgra\ to \sgrb\ and then to us, and the
ISM column density toward \sgrb\ ($N_{\rm H}$). Using Monte Carlo
simulations of the radiative transfer in the gas cloud, we find the
following best-fit values: $\Gamma=1.8\pm0.2$, $\tau=0.4\pm0.1$,
$A=1.9\pm0.2$, $\theta=80^\circ\pm 10^\circ$, and $N_{H}=(8\pm 2)\times
10^{22}$\,cm$^{-2}$ (given are 1$\sigma$ statistical errors). The
uncertainties associated with the gas density 
distribution in \sgrb\ and with the cross-calibration of the
instruments can lead to additional $\sim$30\% systematic uncertainties on
the values of $\tau$ and $\theta$, while our estimates of the iron 
abundance and spectral slope are more robust. The best-fit model is
shown in Fig.~\ref{fig:spectrum}.

Based on the measured optical depth we can estimate the mass of the scattering
gas in \sgrb\ as $M_{{\rm H}_2}=(4\pi/3)\mpr(2n_{{\rm
H}_2})r^3=(4\pi/3)(\mpr/\sigmat)\taut r^2\approx 2\times
10^{6}M_\odot(r/10\,{\rm pc})^2$.

In the above model the high-energy rollover tentatively seen in the
INTEGRAL spectrum is explained by the Compton recoil of hard X-ray
photons in the cloud and is very sensitive to $\theta$. Allowing for
the possibility of an intrinsic cutoff in the illuminating spectrum,
we are able to place an upper limit of $\theta<135^\circ$ on the
mutual position of \sgra\ and the \sgrb\ cloud. 

Using the measured X-ray flux from \sgrb\ and best-fit spectral
parameters, we find that the luminosity of \sgra\ in the 2--10\,keV
and 2--200\,keV band was 0.5 and 1.5$\times
10^{39}$\,erg\,s$^{-1}(r/10\,{\rm pc})^{-2}(d/100\,{\rm pc})^2$,
respectively, where $d$ is the distance between \sgra\ and \sgrb.

\subsection{Alternative scenarios}

Alternative explainations of the X-ray emission of \sgrb\ meet severe
difficulties. Instead of attributing the primary emission to \sgra\ one
could hypothesize that  a transient source inside the \sgrb\ cloud was
irradiating the molecular gas. The large equivalent width of the
6.4\,keV line implies that we are seeing pure reprocessed emission but not
the primary source. The source therefore should have faded away before
the ASCA observations of 1993, i.e. more than 10 years ago. Since the
light crossing time of the \sgrb\ cloud is $\sim 30$ years, one would
expect to see a decline of the 6.4\,keV line flux by a factor of 2
from 1993 till now \citep{sc98}. 

\begin{figure}
\includegraphics[width=\columnwidth,bb=49 182 568 504,clip]{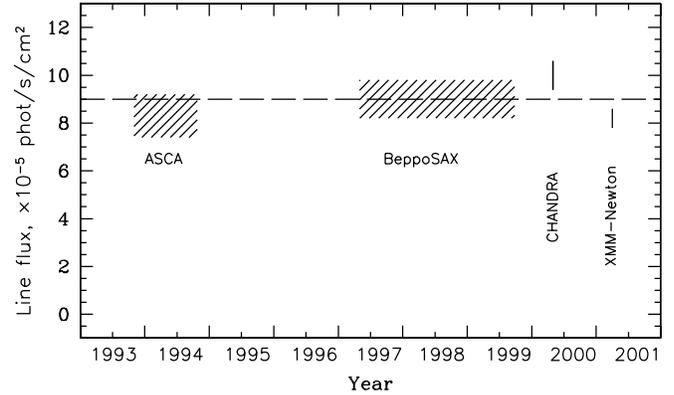}
\caption{Summary of flux measurements in the 6.4\,keV line from the
\sgrb\ cloud. 1$\sigma$ uncertainties are indicated. See text for details.}
\label{fig:lcurve}
\end{figure}

Using archival data of ASCA,
BeppoSAX, Chandra and XMM observatories we find no significant variability
of the line flux during the period 1993--2001 (see
Figure~\ref{fig:lcurve}) in contradiction with the internal
source hypothesis. For each X-ray telescope, we extracted the 6.4\,keV line
flux from the same $3'$ radius circular region centered at the peak
(R.A.=266.830126, Dec.=$-$28.386593) of the Chandra image in the
6.4\,keV line. The background emission (instrumental plus diffuse sky
background unrelated to Sgr B2) was similarly estimated for all the
instruments in a $3'$-radius region around R.A=266.874812,
Dec.=$-$28.501369. The data were reduced using standard utilities
recommended by the Guest Observer Facilities.

INTEGRAL observations similarly indicate that the
continuum 18--60\,keV flux was constant within 25\% during
2003--2004. For the \sgra\ model, the constancy of the line 
flux merely means that the luminosity of \sgra\ remained approximately constant
for more than 10 years a few hundred years ago, while the fact that
other molecular clouds in the GC region also shine in the 6.4\,keV line 
indicates that the entire period of activity lasted much longer than 10
years. The possibility that the GC molecular clouds have been
irradiated by external transient sources such as X-ray binaries has
been ruled out before \citep{sc98,mkm01}. 

 The diffuse emission from the \sgrb\ cloud being a superposition of a
large number of weak point sources inside the cloud is very
unlikely. First of all, the cumulative emission of X-ray sources
(mostly young, low mass stars) in nearby, well-studied molecular
clouds such as Orion and $\rho$ Oph is substantially softer
\citep{rsj+04} than that of \sgrb\ and does not 
exhibit a strong intrinsic 6.4\,keV emission line \citep{fbg+02,ikt01}. The
observed huge equivalent width of the 6.4\,keV line cannot then be
explained by reprocessing of the point sources' emission by
the molecular gas. An even stronger constraint comes from the fact that
several GC molecular clouds with hydrogen column densities ranging between
$10^{23}$ and $10^{24}$\,cm$^{-2}$ \citep{bms+02,pmb+04} all
exhibit a similar equivalent width of the 6.4\,keV line. We briefly note that
for reflecting clouds with a small optical depth, the absorption edge
above 7.1\,keV \citep{pch+03} need not be strong even if the 6.4\,keV line
equivalent width is large.

 The bombardment of molecular gas by low energy cosmic ray electrons
was put forward \citep{vta+00,ylw02} to explain the 6.4\,keV
emission. The electrons produce inner-shell ionizations 
of iron atoms, leading to 6.4\,keV line emission, and
simultaneously generate bremsstrahlung radiation. In this
model, the lack of a strong cutoff below 200\,keV in the spectrum of \sgrb\
implies that electrons with energies higher than a few hundred
keV should be present, while the slope of the observed
spectrum ($\Gamma\approx 2$) constrains the distribution of electrons in
energy. Given these observational constraints, we can estimate in the
thick target approximation that only (1--3)$\times
10^{-5}$ of the cosmic ray electrons' energy can go into hard X-ray
radiation around 50\,keV. Thus, to produce the observed luminosity at
50\,keV of $\sim 3\times 10^{35}$\,erg\,s$^{-1}$ at least (1--3)$\times
10^{40}$\,erg\,s$^{-1}$ of energy in cosmic ray electrons ought to be 
dumped into the cloud. This power is comparable to the bolometric (mostly
infrared) luminosity of \sgrb\ \citep{gbm+93} which is thought to be
mostly due to hot stars. Since \sgrb\ with its dust is an almost 
perfect calorimeter, no room is left for additional energy in
nonthermal, low energy cosmic ray protons. Furthemore, the equivalent
width of the 6.4\,keV line (with respect to the bremsstrahlung 
continuum) is predicted to be 250--350\,eV for solar abundance of
iron. Therefore, the observed $\sim$2\,keV equivalent width requires a
factor of 5--6 overabundance of iron in \sgrb. 

Production of a 6.4\,keV line by cosmic ray ions rather than electrons
\citep{dii+98,trk98} requires similar high energetics and strong 
overabundance of iron. In addition, heavy ions (e.g. oxygen) should
produce multiple ionizations of iron atoms leading to a blue shift of
the 6.4\,keV line, which is not observed. Charge exchange reactions of
nonthermal iron ions with the ambient H$_2$ and He should produce a
broad hump at $\sim$6.7\,keV, which is not observed in \sgrb\
either. The cosmic rays model thus encounters very serious problems.

\section{Conclusion}

We thus come to the conclusion that the \sgrb\ cloud is sending us an
X-ray echo of violent activity of the GC supermassive BH some 300
years ago, which lasted at least 10 years. The luminosity of \sgra\ at
that time was $\approx$5$\times 10^{38}$\,erg\,s$^{-1}$ in the 2--10\,keV
band, i.e. a few $\times 10^5$ times higher than it is now
\citep{bmm+03}. The 2--200\,keV luminosity was $\approx$1.5$\times
10^{39}$\,erg\,s$^{-1}$. \sgra\ was therefore 
similar to low luminosity active galactic nuclei (LLAGN) rather than to
more powerful Seyfert galaxies. LLAGN emit most of their
energy in the near infrared \citep{h99} and we can estimate that the
bolometric luminosity of \sgra\ was $\sim 
10^{40}$\,erg\,s$^{-1}$. This is still only $\sim 3\times 10^{-5}$ of 
the critical Eddington luminosity of the central BH. Thanks to
INTEGRAL, we now know that \sgra\ had a power-law spectrum 
with a photon index of $\sim$1.8, without a significant cutoff up to
100\,keV. This spectrum is remarkably similar 
to the few directly measured hard X-ary spectra of LLAGN
\citep{pcb+00a,pcb+00b}.

It is actually not surprising that our GC was so active in the recent
past, as AGN with luminosities higher than $10^{38}$\,erg\,s$^{-1}$
(2--10\,keV) are found in $\sim$50\% of galaxies 
morphologically similar (of type Sb) to the Milky Way
\citep{hfs97,hft+01}. There is a significant probability that \sgra\
will become bright again in the foreseeable future. That would 
provide the much needed information about the duty cycle of activity 
in galactic nuclei. 

Future instruments will be able to further test the \sgra\ irradiation
model. In particular ASTRO-E2 will probe the Compton shoulder with a
peculiar spectrum on the low-energy side of the 6.4\,keV line, caused
by down-scattering of the line photons on the molecular hydrogen and
helium atoms \citep{sc96,vsc98,suc99}. The relative strength of the
shoulder should be proportional to the optical depth of individual molecular
clouds. In addition, the Compton scattered X-ray continuum should be
more than 50\% polarized in contrast to the fluorescent line
\citep{css02}, and future X-ray polarimeters will be able to determine
the scattering angle and hence the location of \sgrb\ on the line
of sight. Polarization mapping of the GC molecular
clouds would yield a full 3D geometry of the region.

\begin{acknowledgements}
This work is based on observations belonging to the Russian
share (proposal IDs 120213 and 0220133) in the observing time of
INTEGRAL, an ESA project with instruments and science data center
funded by ESA member states (especially the PI countries: Denmark,
France, Germany, Italy, Switzerland, Spain), Czech Republic and
Poland, and with the participation of Russia and the USA. We thank
Kevin Hurley for sharing with us his INTEGRAL TOO observations.  This
research has made use of data obtained through the High Energy
Astrophysics Science Archive Research Center Online Service, provided
by the NASA/Goddard Space Flight Center.
\end{acknowledgements}

\end{document}